\RequirePackage{fix-cm}
\documentclass[twocolumn,final]{svjour3}          % twocolumn
\smartqed  % flush right qed marks, e.g. at end of proof
\usepackage{graphicx}
\usepackage{mathptmx}      % use Times fonts if available on your TeX system
\usepackage{SIunits}
\usepackage{etoolbox}
\usepackage{amsmath}
\usepackage{bm}
\usepackage{mathptmx}
\usepackage[symbol]{footmisc}
\usepackage{microtype}
\usepackage[numbers,sort&compress]{natbib}
\usepackage[colorlinks,allcolors=blue]{hyperref}
\usepackage{textcomp}
\newcommand*{\affaddr}[1]{#1} % No op here. Customize it for different styles.
\newcommand*{\affmark}[1][*]{\textsuperscript{#1}}
 \journalname{}
\begin{document}

\setlength{\abovedisplayskip}{12pt}
\setlength{\belowdisplayskip}{12pt}
\setlength{\abovedisplayshortskip}{6pt}
\setlength{\belowdisplayshortskip}{6pt}

\title{Optical frequency comb Faraday rotation spectroscopy%\thanks{Grants or other notes
}


\author{{Alexandra C. Johansson\affmark[1]\textsuperscript{,$\dagger$}}\and{Jonas Westberg\affmark[2]\textsuperscript{,$\dagger$}}\and{Gerard Wysocki\affmark[2]}\and\mbox{Aleksandra Foltynowicz\affmark[1]}}

\authorrunning{A. C. Johansson et al.} % if too long for running head

\institute{Aleksandra Foltynowicz\\
		\and \email{aleksandra.foltynowicz@umu.se}\\
		\\
		Gerard Wysocki\\
		\and \email{gwysocki@princeton.edu}\\
		\\
		\textsuperscript{$\dagger$}These authors contributed equally to this work.\\
		\\
		\affaddr{\affmark[1]Department of Physics}\\
		\affaddr{Ume\aa\, University}\\
		\affaddr{901 87 Ume\aa\,, Sweden}\\
		\\
		\affaddr{\affmark[2]Department of Electrical Engineering}\\
		\affaddr{Princeton University}\\
		\affaddr{Princeton, New Jersey 08544, USA}
                         %  \\
}

\date{Received: 25 Janauray 2018 }

\maketitle

%%%%%%%%%%%%%%%%%
%%%%%  ABSTRACT%%%%%%%
%%%%%%%%%%%%%%%%%

\begin{abstract}
We demonstrate optical frequency comb Faraday rotation spectroscopy (OFC-FRS) for broadband interference-free detection of paramagnetic species. The system is based on a femtosecond doubly resonant optical parametric oscillator and a fast-scanning Fourier transform spectrometer (FTS). The sample is placed in a DC magnetic field parallel to the light propagation. Efficient background suppression is implemented via switching the direction of the field on consecutive FTS scans and subtracting the consecutive spectra, which enables long term averaging. In this first demonstration, we measure the entire Q- and R-branches of the fundamental band of nitric oxide in the 5.2-\unit{5.4}{\micro\meter} range and achieve good agreement with a theoretical model.
\end{abstract}

%%%%%%%%%%%%%%%%%
%%%%%   INTRO     %%%%%%%
%%%%%%%%%%%%%%%%%

\section{Introduction}
\label{intro}
Laser-based spectroscopy for quantitative gas detection offers high sensitivity, high species specificity and the possibility of performing real-time measurements in situ without sample consumption. Laser-based gas sensors have found applications in fields ranging from fundamental science \cite{truong_accurate_2015} and environmental monitoring \cite{curl_quantum_2010,hodgkinson_optical_2013,wang_optical_2014} to medical diagnostics \cite{wang_breath_2009,svanberg_diode_2016} and security \cite{todd_application_2002,li_contributed_2015}. However, the inherently limited optical bandwidth of these sensors has mostly restricted them to single species detection in spectral regions that are free from spectral interferences. The development of optical frequency combs has largely addressed this issue by massively extending the attainable spectral bandwidth, while retaining the high resolution \cite{diddams_molecular_2007}. Contemporary, commercially available, frequency comb systems now span hundreds of wavenumbers with optical power distributed across tens of thousands of equidistant comb lines. These sources have opened up for broadband, sensitive, and high-resolution spectroscopy, which allows retrieval of immense amounts of spectroscopic information in short acquisition times \cite{fleisher_mid-infrared_2014,spaun_continuous_2016} and makes multi-species detection with a single coherent light source feasible \cite{adler_mid-infrared_2010, grilli_frequency_2012,rieker_frequency-comb-based_2014}. Computational methods involving baseline removal together with multi-line fitting can be straightforwardly implemented to retrieve quantitative information from these multi-species spectra if accurate spectroscopic models of each species are known \cite{adler_mid-infrared_2010}. However, quantifying the concentrations of each constituent in a congested absorption spectrum is a challenging task when sufficiently accurate spectral models for the background constituents are unknown and their spectra cannot be assigned. The difficulties may be further exacerbated by drifting optical fringe backgrounds that prohibit long-term averaging. Congested absorption spectra are commonly encountered in e.g. combustion diagnostics, and are especially troublesome in wavelength regions where sufficiently detailed spectral models for water vapor are not yet available \cite{rutkowski_detection_2016}.

In case of paramagnetic species any potential spectral interference can be effectively suppressed by implementing the interference- and background-free Faraday rotation spectroscopy (FRS) technique \cite{litfin_sensitivity_1980,blake_prognosis_1996,lewicki_ultrasensitive_2009}, which relies on probing the opto-magnetic properties of paramagnetic species subjected to an external magnetic field. In this work, we merge the spectral bandwidth of an optical frequency combs with FRS to obtain a broadband high-resolution spectroscopy system that can provide simultaneous access to the entire ro-vibrational band of the target paramagnetic species. By placing the sample inside a solenoid (providing the external magnetic field) located between two polarizers, the polarization rotation induced by the molecules is translated into an intensity change that is directly proportional to the molecular number density. The direction and magnitude of the magnetic field can be controlled through the current of the solenoid, which allows for direct modulation of the Faraday rotation signal. The insusceptibility of diamagnetic species, such as H\textsubscript{2}O and CO\textsubscript{2}, to the Faraday effect infers that interferences from these species are efficiently suppressed (with the caveat that a sufficient number of photons still reach the photodetector). The contributions from the paramagnetic species can therefore be disentangled from a congested absorption spectrum, which significantly simplifies the spectroscopic assignments. Magnetic field modulation also eliminates the transmission baseline, thereby allowing prolonged averaging times even outside a temperature controlled laboratory environment. 

In this first demonstration of optical frequency comb Faraday rotation spectroscopy (OFC-FRS) we detect interference-free spectra of nitric oxide (NO) using a system based on a femtosecond doubly resonant optical parametric oscillator (DROPO) and a fast-scanning Fourier transform spectrometer (FTS). We measure the spectrum of the entire Q- and R-branches of NO at 5.2-\unit{5.4}{\micro\meter}, and the measurement shows good agreement with a theoretical model.
%%%%%%%%%%%%%%%%%
%%%%%  THEORY  %%%%%%%
%%%%%%%%%%%%%%%%%
\raggedbottom
\section{Theory}
\label{sec:theory}
\begin{figure}
  \includegraphics[width=\columnwidth]{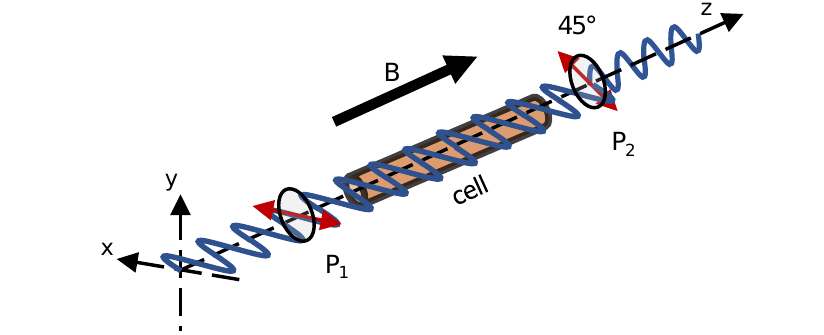}
\caption{Schematic overview of the electric field propagation in an FRS system. P\textsubscript{1}, polarizer; P\textsubscript{2}, analyzer; B, magnetic field vector. }
\label{fig:1}  
\vspace{-12pt}     % Give a unique label
\end{figure}

A general overview of the light propagation in an FRS system with DC magnetic field is shown in Fig. \ref{fig:1}. The polarization of the incoming light is oriented horizontally ($x$-axis) as indicated by the red arrow. The magnetically induced polarization rotation due to the paramagnetic absorber is converted to an intensity change using a polarizer (analyzer) after the sample cell. The choice of the analyzer angle depends on the noise properties of the system, where small angles are used for systems with large relative intensity noise and \unit{45}{\degree} is used for a detector noise limited system \cite{lewicki_ultrasensitive_2009}.

Consider a horizontally polarized (x-direction) electric field impinging on the sample, where a small ellipticity, $\varepsilon$ , has been added to account for the finite extinction ratio of the polarizer \cite{brecha_analysis_1999,westberg_lineshape_2014}. The electric field before the interaction with the sample can be written in Jones’ notation \cite{jones_new_1948} as
\begin{align}
\label{eq:1}
	\bm{E}_i = E_0 \begin{bmatrix} 1 \\ i\varepsilon \end{bmatrix}e^{i(kz-\omega t)},
\end{align}
where $E_0$ is the electric field amplitude, $k$ the wave vector, $\omega$ the angular frequency, and $t$ the time. This expression can be rewritten as a sum of the left-hand (LCP) and right-hand (RCP) circular components expressed in terms of their unit vectors
\begin{align}
	\label{eq:2}
	&\hat{e}_L = \frac{1}{\sqrt{2}}\begin{bmatrix} 1 \\ i \end{bmatrix}, & \hat{e}_R = \frac{1}{\sqrt{2}}\begin{bmatrix} 1 \\ -i \end{bmatrix},
\end{align}
which gives,
\begin{align}
	\label{eq:3}
	\bm{E}_i  = \frac{E_0}{\sqrt{2}}e^{ik_0L-\omega t}\left[\hat{e}_L\left(1-\varepsilon\right)e^{-\delta_L + i\phi_L} + \hat{e}_R\left(1+\varepsilon\right)e^{-\delta_R + i\phi_R}\right], 
\end{align}
where $L$ is the interaction length, $k_0$ is the wave vector in the absence of absorbers, and  $\delta_{L,R}$ and $\phi_{L,R}$  denotes the attenuation and phase shift experienced by the circular components of the electric field upon interaction with the absorber. $\delta_{L,R}$ and $\phi_{L,R}$ are related to the extinction coefficient, $\kappa$, and the refractive index, $n$, through
\begin{align}
	\label{eq:4}
	& \delta_{L,R} = \kappa_{L,R}\frac{\omega L}{c}, & \phi_{L,R} = \left(n_{L,R}-1\right)\frac{\omega L}{c}.
\end{align}
After interaction with the paramagnetic sample subjected to a magnetic field oriented along the light propagation axis, the light is transmitted through a second polarizer (analyzer) oriented at an uncrossing angle, $\theta$, with respect to the first polarizer. The Jones’ matrix for this element is given by \cite{westberg_faraday_2013}
\begin{align}
\label{eq:5}
\bm{J}_{\!pol} =
\!\begin{aligned}
&
\left[\begin{matrix}
  a\sin^2\left(\theta\right) + b\cos^2\left(\theta\right)\\
 a\sin\left(\theta\right)\cos\left(\theta\right)-b\sin\left(\theta\right)\cos\left(\theta\right)
\end{matrix}\right.\\
&\qquad\qquad
\left.\begin{matrix}
  {} a\sin\left(\theta\right)\cos\left(\theta\right)-b\sin\left(\theta\right)\cos\left(\theta\right) \\
  {}a\cos^2\left(\theta\right) + b\sin^2\left(\theta\right)
\end{matrix}\right],
\end{aligned}
\end{align}
where $a$ and $b$ denote the fractions of the electric field transmitted along the main and orthogonal polarization axis of the analyzer. For a high extinction ratio analyzer, $a\approx 1$ and $b \ll 1$. Finally, the transmitted intensity, $I_t$, is given by the product of the transmitted electric field and its complex conjugate, which can be calculated from Eqs. (\ref{eq:3}) and (\ref{eq:5}). This gives
\begin{align}
	\label{eq:6}
	I_t &= \frac{I_0}{2}e^{-2\bar{\delta}} \big\{\!\!\left(a^2+\varepsilon^2b^2\right)\left[\cosh\left(\Delta\delta\right) - \cos\left(\Delta\phi + 2\theta\right)\right]\nonumber \\ 
		&+ 2\left(a^2 + b^2\right)\varepsilon\sinh\left(\Delta\delta\right)\nonumber \\
		&+ \left(\varepsilon^2 a^2 + b^2\right)\left[\cosh\left(\Delta\delta\right) + \cos\left(\Delta\phi+2\theta\right)\right]\!\!\big\},
\end{align}
where $\bar{\delta}$ is the average of the attenuations experienced by the LCP and RCP light, $\Delta\delta$ is the differential attenuation due to magnetic circular dichroism (MCD), and $\Delta\phi$ is the differential phase shift due to magnetic circular birefringence (MCB). These entities are in turn given by
\begin{align}
	\label{eq:7}
	2\bar{\delta} &= \delta_L + \delta_R = \frac{\alpha_0L}{2}\left(\chi^{abs}_{L} + \chi^{abs}_{R}\right) = \bar{\alpha} L \nonumber \\ 
	\Delta\delta &= \delta_L - \delta_R = \frac{\alpha_0L}{2}\left(\chi^{abs}_{L} - \chi^{abs}_{R}\right) = 2\Phi \nonumber \\
	\Delta\phi &= \phi_L - \phi_R = \frac{\alpha_0L}{2}\left(\chi^{disp}_{L} - \chi^{disp}_{R}\right) = 2\Theta, 	
\end{align}
where $\alpha_0$ is the on-resonance absorption coefficient given by $\alpha_0 = S'c_{rel}p$, where $S'$ is the integrated gas linestrength (cm\textsuperscript{-2}/atm)%\footnote[2]{$S'$ is related to the HITRAN integrated linestrength, $S$, through $S' = 2.6868\times 10^{19} \left(T_0/T\right) S$, where $T_0$ is the reference temperature (K) and  $T$ the measurement temperature (K).}
, $c_{rel}$ is the relative concentration, $p$ the total gas pressure (atm), and $\bar{\alpha}$ the average absorption coefficient. $ \chi^{abs,disp}_{L,R}$ are the absorption and dispersion lineshape functions for the LCP and RCP components given by \cite{westberg_faraday_2013}
\begin{align}
\label{eq:8}
\chi^{abs,disp}_{L,R} = 3\sum_{M''_JM'_J} \begin{pmatrix} J' & 1 & J''\\-M'_J & \Delta M_J & M''_J \end{pmatrix}^2 \chi_V^{abs,disp}(\nu_{M''_JM'_J}),
\end{align}
where the $\sum(...)^2$ term is the Wigner 3-$j$ symbol for a $M''_J \rightarrow M'_J$ transition, and $\Delta M_J =  M'_J - M''_J$ (equal to +1 or -1 for LCP and RCP, respectively). $\chi_V^{abs,disp}(\nu_{M''_JM'_J})$ denotes the Voigt absorption and dispersion lineshapes and $\nu_{M''_JM'_J}$  is the detuning frequency shifted by the external magnetic field, given by
\begin{align}
\label{eq:9}
\nu_{M''_JM'_J} = \left[\nu - \nu_0 - \left(M'_Jg'_J - M''_Jg''_J\right)\mu_B B\right]\sqrt{\ln(2)}/\delta\nu_D,
\end{align}
where $\nu$ is the frequency of the light, $\nu_0$ is the unperturbed molecular transition frequency, $g_J$ the Land\'e g-factor, $\mu_B$ the Bohr magneton, $B$ the external magnetic field, and $\delta\nu_D$ the Doppler width (FWHM) of the transition. 

The Faraday rotation angle, $\Theta$, is defined as half of the induced differential phase shift, i.e. $\Theta = \Delta\phi/2$ [see Eq. (\ref{eq:7})]. Since the sign of $\Theta$ depends on the direction of the magnetic field \cite{westberg_quantitative_2010}, efficient background suppression can be achieved by alternating the direction of the magnetic field and subtracting the spectra measured with opposite fields. Using Eq. (\ref{eq:6}), the transmitted intensities for the two field directions, denoted $I_t^{+}$ and $I_t^{-}$, where the $+$ and $-$ signs indicate the direction of the magnetic field, can be written as
\begin{align}
	\label{eq:10}
	I_t^{\pm} &= \frac{I_0}{2}e^{-\bar{\alpha} L} \big\{\!\!\left(a^2+\varepsilon^2b^2\right)\left[\cosh\left(\pm 2\Phi\right) - \cos\left(\pm 2\Theta + 2\theta\right)\right]\nonumber \\ 
		&+ 2\left(a^2 + b^2\right)\varepsilon\sinh\left(\pm 2\Phi\right)\nonumber \\
		&+ \left(\varepsilon^2 a^2 + b^2\right)\left[\cosh\left(\pm 2\Phi\right) + \cos\left(\pm 2\Theta+2\theta\right)\right]\!\!\big\}.
\end{align}
It follows that their average, $I_t^{0} = (I_t^{+}+I_t^{-})/2$, is given by
\begin{align}
	\label{eq:11}
	I_t^{0} &= \frac{I_0}{4}e^{-\bar{\alpha} L} \big\{\!\!\left(a^2+\varepsilon^2b^2\right)\left[2\cosh\left(2\Phi\right) - 2\cos\left(2\theta\right) \cos\left(2\Theta\right)\right] \nonumber \\ 
		&+ \left(\varepsilon^2 a^2 + b^2\right)\left[2\cosh\left(2\Phi\right) + 2\cos\left(2\theta\right)\cos\left(2\Theta\right)\right]\!\!\big\},
\end{align}
which for small $\Phi$ and high extinction ratio polarizers $(\varepsilon^2b^2 \ll \varepsilon^2a^2 \approx b^2 \ll a^2 \simeq 1)$ reduces to
\begin{align}
	\label{eq:12}
	 I_t^{0} &= \frac{I_0}{2}e^{-\bar{\alpha} L} \left[1 - \cos\left(2\theta\right)\cos\left(2\Theta\right) \right].
\end{align}
When the polarizer uncrossing angle $\theta$ is set to \unit{45}{\degree}, this simply gives the Beer-Lambert law with half of the power transmitted through the analyzer. The difference of the transmitted intensities measured with opposite magnetic field directions, $\Delta I_t^{\pm} = I_t^{+} - I_t^{-}$, is given by
\begin{align}
	\label{eq:13}
	\Delta I_t^{\pm} &= I_0 e^{-\bar{\alpha} L} \big\{\!\!\sin\left(2\theta\right)\sin\left(2\Theta\right) \left[\left(a^2+\varepsilon^2b^2\right) -  \left(\varepsilon^2 a^2 + b^2\right)\right] \nonumber \\
	&+2\left(a^2 + b^2\right)\varepsilon\sinh\left(2\Phi\right)\!\!\big\},
\end{align}
which for high extinction ratio polarizers reduces to 
\begin{align}
	\label{eq:14}
	\Delta I_t^{\pm} &= I_0 e^{-\bar{\alpha} L} \left[ \sin\left(2\theta\right)\sin\left(2\Theta\right) +2\varepsilon\sinh\left(2\Phi\right)\right].
\end{align}
A normalized FRS signal can be obtained by taking the ratio of the differential intensities and their mean, i.e. 
\begin{align}
	\label{eq:15}
	S_{FRS} =\frac{\Delta I_t^{\pm}}{I_t^{0}} \approx  \frac{2\sin\left(2\theta\right)\sin\left(2\Theta\right) +4\varepsilon\sinh\left(2\Phi\right)}{1-\cos\left(2\theta\right)\cos\left(2\Theta\right)},
\end{align}
where the last equality assumes high extinction ratio polarizers. The MCD contribution $\Phi$ [see Eq. (\ref{eq:7})] is coupled through the extinction ratio of the first polarizer, $\varepsilon$, which introduces an asymmetry in the lineshape \cite{brecha_analysis_1999,westberg_lineshape_2014}. Note that the dependence on $I_0e^{-\bar{\alpha} L}$ is removed by the normalization procedure of Eq. (\ref{eq:15}), which infers that any variations in intensity that are common for the two signals are suppressed. This includes interferences from diamagnetic species, such as water vapor, and from slowly varying optical fringes affecting the transmission baseline.

For the special case of a system limited by detector noise, the highest signal-to-noise ratio is obtained when the uncrossing angle is set to \unit{45}{\degree}  \cite{blake_prognosis_1996,lewicki_ultrasensitive_2009}. This further simplifies the expression above to
\begin{align}
	\label{eq:16}
	S_{FRS} = 2\sin\left(2\Theta\right) + 4\varepsilon\sinh\left(2\Phi\right).
\end{align}
For small rotation angles and high extinction ratio polarizers the normalized FRS signal is directly proportional to the Faraday rotation angle, which can be used to extract the sample concentration \cite{westberg_quantitative_2010}. For the cases where the aforementioned approximations are not justified, the full expressions of Eqs. (\ref{eq:11}) and (\ref{eq:13}) must be used to model the FRS spectrum.
\raggedbottom

%%%%%%%%%%%%%%%%%
%%%%%  EXPERIMENTAL  %%%%
%%%%%%%%%%%%%%%%%
\section{Experimental setup and procedures}
\label{sec:exp}
\begin{figure}
  \includegraphics[width=\columnwidth]{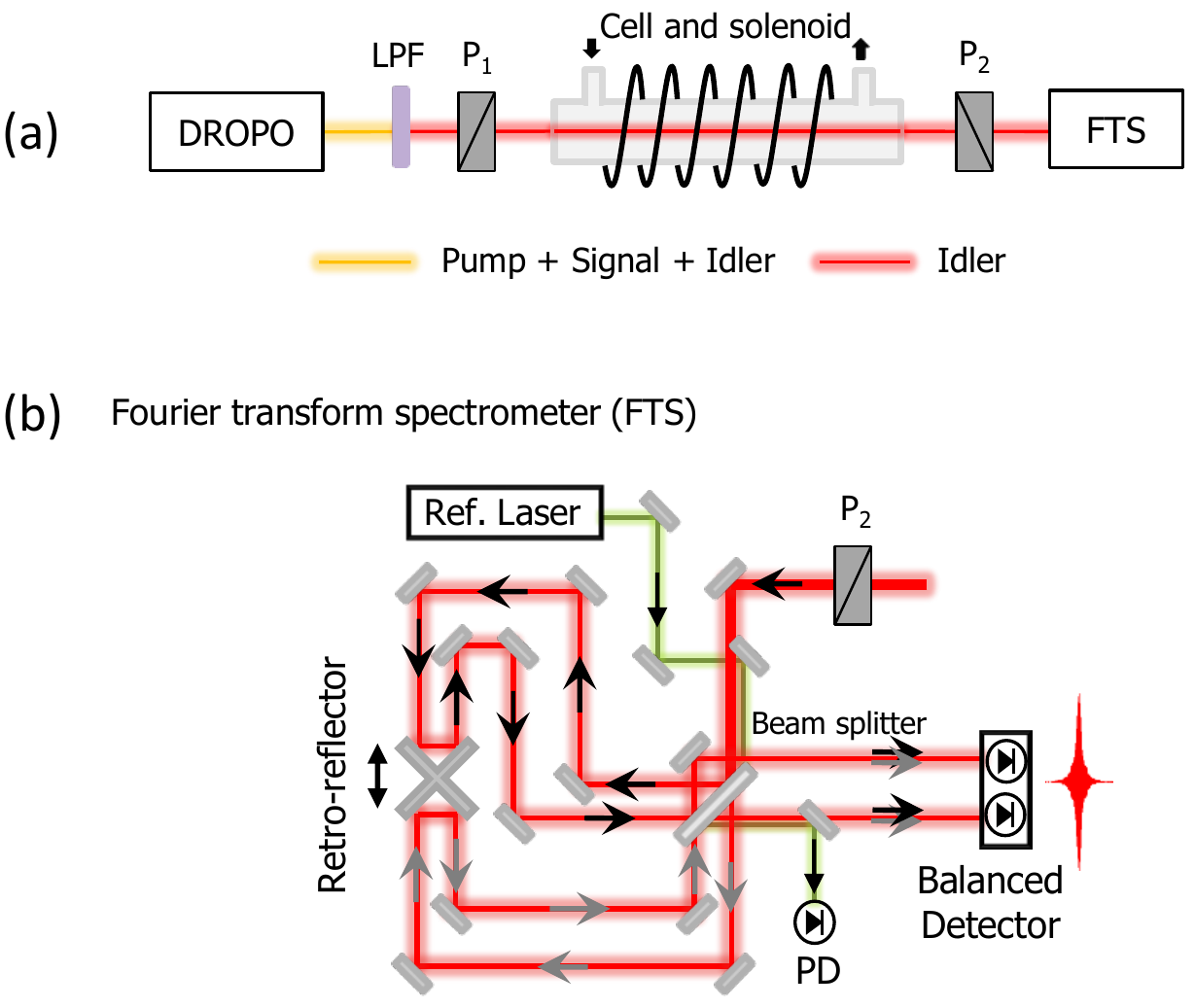}
\caption{(a) Experimental setup. DROPO, doubly resonant optical parametric oscillator; LPF, long-pass filter; P\textsubscript{1}, Wollaston prism polarizer, P\textsubscript{2}, wire-grid polarizer; FTS, Fourier transform spectrometer. (b) Fourier transform spectrometer. PD; photodetector.}
\label{fig:2}       % Give a unique label
\vspace{-12pt}
\end{figure}

The experimental setup, shown in Fig. \ref{fig:2}(a), is based on a femtosecond doubly resonant optical parametric oscillator (DROPO), two polarizers, a gas cell placed inside a DC-solenoid, and a fast-scanning Fourier transform spectrometer (FTS). A detailed description of the light source is provided in Ref. \cite{khodabakhsh_fourier_2016}. The DROPO is synchronously pumped with a mode-locked Tm:fiber laser with a repetition rate of \unit{125}{\mega\hertz}. The idler of the DROPO operating in the degenerate configuration is tuned to cover 5.1-\unit{5.4}{\micro\meter} (1850-\unit{1960}{\centi\reciprocal\meter}) with a total optical power of \unit{5}{\milli\watt} by adjusting the length of the DROPO cavity. A long-pass filter after the DROPO is used to filter out the pump and signal light. The polarization of the idler light is cleaned by a Wollaston prism placed in front of a \unit{17.5}{\centi\meter}-long gas cell equipped with uncoated, tilted, and wedged CaF$_2$ windows. The cell is placed inside a \unit{12.7}{\centi\meter}-long DC-solenoid, which provides an axial magnetic field of 260$\pm$\unit{7}{G} either parallel or antiparallel to the light propagation depending on the direction of the supplied current. The cell is operated with a continuous flow of 1$\%$ NO diluted in N$_2$ at a total pressure of \unit{100}{Torr}. The flow and pressure are actively controlled through a combination of flow and pressure controllers (providing \unit{1.3}{Torr} precision on the gas pressure). The magnetic field strength was determined from the intensity of the current supplied to the solenoid using a calibration curve obtained earlier by direct measurements with a Gaussmeter. A wire-grid polarizer, with the $a$ and $b$ parameters measured to be $a = 0.67$, and $b = 0.40$, is placed after the cell and rotated by \unit{45}{\degree} with respect to the Wollaston prism to convert the polarization rotation induced by the interaction with the sample to an intensity change. The transmitted light is measured with a fast-scanning FTS, 
illustrated in Fig. 2(b), equipped with two HgCdTe detectors in balanced configuration \cite{khodabakhsh_fourier_2016}. The two out-of-phase outputs from the FTS are subtracted, which allows for common-mode noise suppression (and higher signal-to-noise ratios). An interferogram with a spectral resolution of \unit{250}{\mega\hertz} is measured in \unit{3.5}{\second} and a stable continuous-wave laser at \unit{1.56}{\micro\meter} is used for optical path difference calibration. Consecutive interferograms are collected with opposite magnetic field by synchronizing the direction of the supplied solenoid current with the movement of the FTS mirrors. The corresponding spectra, $I_t^{\pm}$, are retrieved from fast Fourier transform of the interferograms. The normalized FRS spectrum is obtained by subtracting spectra recorded with opposite magnetic field, $\Delta I_t^{\pm} = I_t^{+} - I_t^{-}$, and normalizing to their mean value, $I_t^{0} = \left(I_t^{+} + I_t^{-}\right)/2$, which makes the FRS spectrum baseline- and calibration-free.

%%%%%%%%%%%%%%%%%
%%%%%  RESULTS  %%%%
%%%%%%%%%%%%%%%%%
\section{Results}
\label{sec:res}
\begin{figure}
  \includegraphics[width=\columnwidth]{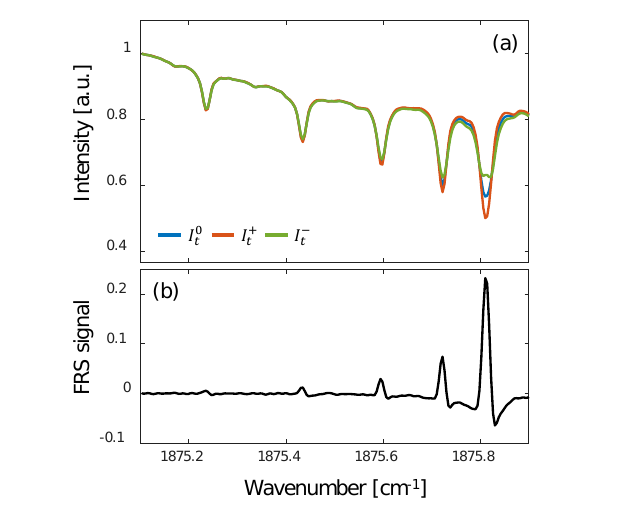}
\caption{(a) Transmission spectra after Fourier transform of the interferograms (500 averages) for opposite magnetic field directions ($I_t^{\pm}$, red and green) and their mean ($I_t^0$, blue). (b) The corresponding normalized FRS spectrum.}
\label{fig:3}       % Give a unique label
\vspace{-12pt}
\end{figure}

Figure \ref{fig:3}(a) shows an average of 500 transmission spectra measured with opposite magnetic field, $I_t^{+}$ in red and $I_t^{-}$ in green, and their mean, $I_t^0$, in blue, zoomed in on the $\mbox{X}^2\mathrm{\Pi}_{3/2}(\nu=1\rightarrow 0)$ Q-branch of NO at \unit{1875.8}{\centi\reciprocal\meter}. The rotation of the polarization of the light due to the magnetic field is manifested as either an increase or a decrease of intensity depending on the direction of the field intensity (compared to the case with no magnetic field, which corresponds to the mean value). Figure 3(b) shows the resulting normalized FRS spectrum. Subtraction of the consecutive transmission spectra removes any structure in the baseline, which can clearly be seen by comparing the upper and lower panels of Fig. \ref{fig:3}. 

\begin{figure}
  \includegraphics[width=\columnwidth]{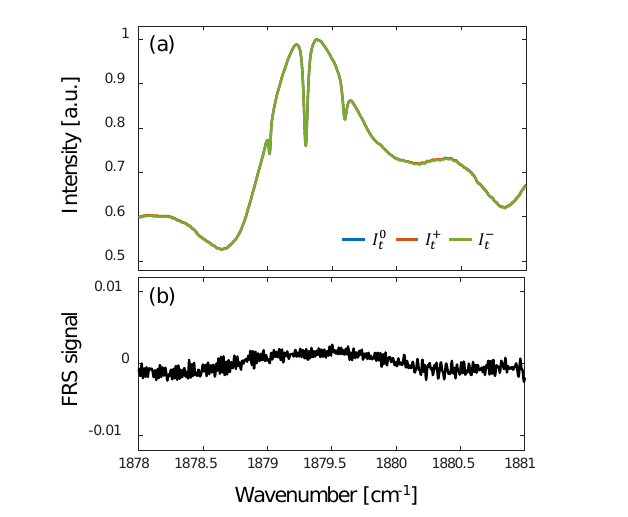}
\caption{(a) Transmission spectra (500 averages) at a wavelength range affected by water absorption. The structure on the baseline is due to optical fringes introduced by the system optics. (b) The corresponding normalized FRS spectrum, where water absorption and etalons are largely suppressed.}
\label{fig:4}       % Give a unique label
\vspace{-12pt}
\end{figure}

Figure \ref{fig:4} further demonstrates the water and optical fringe suppression capabilities, where (a) displays a part of the transmission spectrum (500 averages) and (b) shows the resulting normalized FRS spectrum. This region is visibly affected by optical fringes, originating from the analyzer, and by water absorption, caused by the presence of atmospheric water in the unpurged beam path between the DROPO and the FTS. The water absorption reaches almost 20$\%$ for the strongest transition. However, the diamagnetic water molecules do not induce any polarization rotation and hence two consecutive transmission spectra (with opposing magnetic fields) exhibit nearly perfect overlap as shown in Fig. \ref{fig:4}(a). Figure \ref{fig:4}(b) shows the corresponding normalized FRS spectrum, in which the water absorption and optical fringes are efficiently cancelled. The noise on the baseline over this range is $\sim 5\times10^{-4}$, and is primarily limited by the detector noise. The normalized FRS spectrum does not show any significant structure remaining from the water absorption, however there is a small residual baseline caused by optical fringes that do not cancel out completely due to baseline drift that occurs on a time scale faster than that required to perform two FTS scans. The fluctuation of the baseline caused by these residual fringes is $\sim 2\times 10^{-3}$, which corresponds to a reduction of optical fringes observed in the transmission spectrum by a factor of $\sim 200$. It should also be noted that the measurement is possible only if sufficiently many photons reach the photodetectors. Therefore, in spectral regions where water absorption approaches 100$\%$ the system will exhibit a severe degradation of the signal-to-noise ratio. In this case, water vapor removal by optical path purging is still recommended although the demand on its efficiency is relaxed. 
 
\begin{figure}
  \includegraphics[width=1\columnwidth]{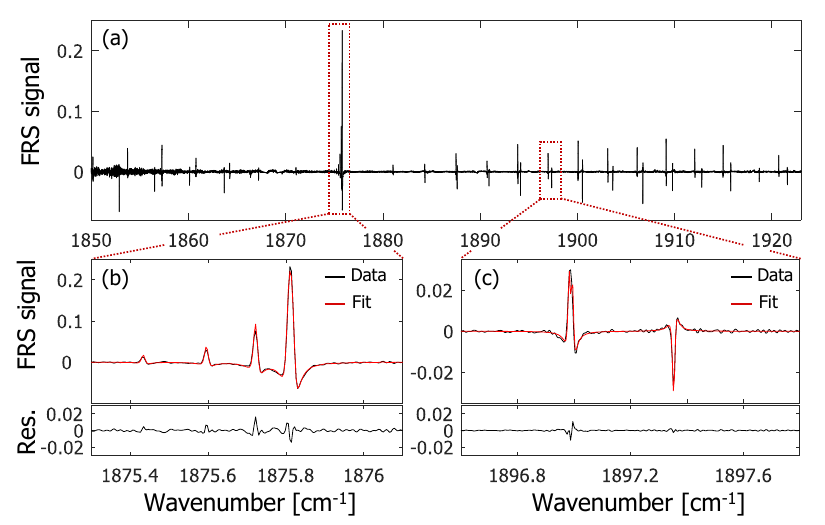}
\caption{(a) Normalized FRS spectrum of the entire Q- and R-branch and part of the P-branch of nitric oxide at $\sim$\unit{5.3}{\micro\meter}. The spectral regions outlined in red are further enlarged in (b) and (c). (b) A zoomed view of the Q-branch (black) at $\sim$\unit{1876}{\centi\reciprocal\meter} together with a fit (red) of a model based on Eqs. (\ref{eq:11}) and (\ref{eq:13}). (c) A close-up of a pair of R-branch transitions (black) at $\sim$\unit{1897}{\centi\reciprocal\meter} together with a fit (red). Residuals from the fits are shown in the lower panels in (b) and (c).}
\label{fig:5}       % Give a unique label
\vspace{-12pt}
\end{figure}

Figure \ref{fig:5}(a) shows the normalized FRS spectrum of 1$\%$ NO in N$_2$ at \unit{100}{Torr}, covering 1850-\unit{1930}{\centi\reciprocal\meter} (500 averages). The residual baseline caused by uncorrelated structure between two transmission spectra has been corrected in the post-processing, and water lines with 100$\%$ absorption have been masked out. Figure \ref{fig:5}(b) shows a zoomed view of the normalized FRS spectrum for the Q-branch (black) together with a fit (red). The strongest normalized FRS signal comes from the Q(3/2) transition of the $\mbox{X}^2\mathrm{\Pi}_{3/2}$ subsystem, and higher J-quantum numbers yield diminishing signal strengths. Figure \ref{fig:5}(c) displays the normalized FRS spectrum (black) for two lines in the R-branch, the R(13/2) and the R(15/2) transitions of the $\mbox{X}^2\mathrm{\Pi}_{1/2}$ and $\mbox{X}^2\mathrm{\Pi}_{1/2}$ subsystems, respectively, together with a fit (red).
 
The fitted model is based on the full expressions of Eqs. (\ref{eq:11}) and (\ref{eq:13}), to account for imperfect polarizers \cite{brecha_analysis_1999}. The spectral line parameters are taken from the HITRAN database \cite{gordon_hitran2016_2017} and the Land\'e g-factors are calculated from Refs. \cite{radford_microwave_1961,brown_l-uncoupling_1966,herrmann_line_1980}. For the Q-branch in Fig. \ref{fig:5}(b) the fitting parameters are the NO concentration, $c_{NO}$, the magnetic field strength, $B$, and the unbalancing term between the right-hand and the left-hand circularly polarized components for the first polarizer, $\varepsilon$. The values returned from the fit are $c_{NO} = 1.055(2 )\%$, B = \unit{228.8(5)}{G}, and $\varepsilon = 0.0507(3)$. The discrepancy from the measured peak field intensity of \unit{260}{G} is likely due to a non-uniform field distribution over the interaction length, and the frequency dependence of the polarizer parameters $a$ and $b$ that is not taken into account. For the fit to the R-branch, $B$ and $\varepsilon$ were fixed to the values returned from the fit to the Q-branch and only the NO concentration was used as a fitting parameter. Here, the fit returned $c_{NO} = 0.944(4)\%$. The residuals in Figs \ref{fig:5}(b) and (c) indicate that the model agrees well with the measured data, although some discrepancies can be observed. These are likely caused by experimental limitations (poor extinction ratio of the wire-grid analyzer and non-linearities in the FTS detectors), as well as imperfections in the theoretical model (i.e. the use of the Voigt lineshape function and uncertainties in the values of the Land\'e g-factors \cite{radford_microwave_1961,brown_l-uncoupling_1966,herrmann_line_1980}).

It should be noted that the optimum magnetic field strength (i.e. the one that maximizes the signal-to-noise ratio) varies for the different branches due to their different Land\'e g-factors. For the NO spectrum, the field strength needed to maximize the signal from the R-branch is higher than that needed to maximize the signal from the Q-branch. The magnitude of the magnetic field used in the experiment was set to the technical maximum that could be sustained for long measurement durations (limited by the heat dissipation of the coil). At this field, the R-branch transitions were strongly undermodulated, while the Q-branch transitions were slightly overmodulated. In general, the magnetic field and sample gas pressure should be optimized for each FRS measurement by taking into account the difference in susceptibility to the magnetic field of the branches, magnetic field limitations, and the noise sources in the system. 

\begin{figure}
  \includegraphics[width=\columnwidth]{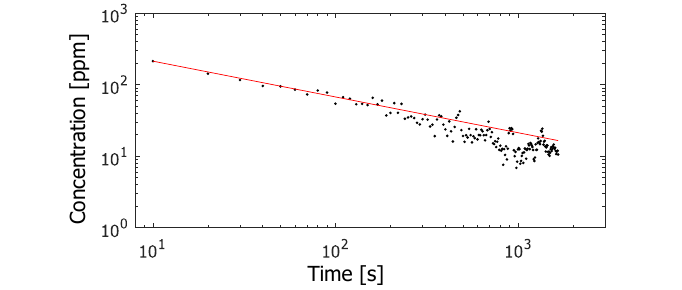}
\caption{Allan-Werle plot of the relative concentration of NO retrieved from the multiline fit to the normalized FRS spectra of the Q-branch (black markers).The red line indicates the white-noise behavior of the system for more than 1000 seconds.}
\label{fig:6}       % Give a unique label
\vspace{-12pt}
\end{figure}

The stability of the system is characterized by an Allan-Werle plot, where the standard deviation of the fitted concentration as a function of averaging time is calculated. The concentrations are extracted from multiline fits to consecutive individual FRS spectra of the Q-branch, with a model based on Eqs. (\ref{eq:11}) and (\ref{eq:13}). The resulting Allan-Werle plot is shown in Fig. \ref{fig:6}, and indicates a white-noise limited behavior for the full durations of the measurement ($>$1000 seconds). As mentioned previously, this is not unexpected given the fact that the normalized FRS spectrum is based on the difference between two consecutive transmission spectra, which is equivalent to time-multiplexed differential detection, albeit with long acquisition times. Based on the Allan deviation presented in Fig. \ref{fig:6}, the $1\sigma$ minimum concentration detection limit of the system was estimated to \unit{1.4}{ppm}$\cdot$m after 1000 seconds of integration time.
% 

%%%%%%%%%%%%%%%%%
%%%%%  CONCLUSIONS  %%%%
%%%%%%%%%%%%%%%%%
\section{Conclusions}
\label{sec:conclusions}
In conclusion, we presented for the first time optical frequency comb Faraday rotation spectroscopy (OFC-FRS) by measuring the spectrum of the entire Q- and R-branches of the fundamental vibrational band of NO at $\sim$\unit{5.3}{\micro\meter} using a femtosecond DROPO and a fast-scanning FTS. Switching the direction of the magnetic field for consecutive FTS scans and subtracting the resulting transmission spectra enables efficient background suppression and allows measurements over long averaging times. Moreover, the normalization by the mean of consecutive spectra provides a calibration-free signal. A theoretical model of the normalized FRS spectrum is presented, which shows good agreement with the measurements. A minimum detection limit for NO of \unit{1.4}{ppm}$\cdot$m at \unit{1000}{\second} is achieved, and the system remains stable for more than 1000 seconds. Further improvements in sensitivity can be obtained by improving the extinction ratio of the polarizers or increasing the interaction length through multiple passes \cite{zhang_hybrid_2014} or by cavity enhancement \cite{westberg_cavity_2017,gianella_intracavity_2017}. While in this first demonstration we used an FTS to acquire broadband OFC-FRS spectra, the technique is compatible with other detection schemes of comb spectroscopy, for example with continuous Vernier filtering \cite{rutkowski_broadband_2014,khodabakhsh_mid-infrared_2017}, which should allow faster acquisition and thus faster modulation of the magnetic field. 

The broadband OFC-FRS technique may be useful for fundamental science applications, such as assessing Land\'e g-factors, where the calibrated frequency axis and immunity to instrumental lineshapes \cite{maslowski_surpassing_2016} are advantages compared to FRS based on tunable cw lasers in the mid-infrared \cite{wu_high-resolution_2017}. The OFC-FRS technique is also applicable when targeting trace amounts of gaseous paramagnetic species masked by excessive quantities of interfering diamagnetic compounds (H$_2$O, CO$_2$, etc.). Such conditions are common in combustion diagnostics, where the large optical bandwidth provided by the comb in combination with the Faraday rotation technique will provide a new tool for probing high temperature chemical reactions involving magnetically sensitive species, e.g. free radicals such as NO, NO$_2$, OH and HO$_2$. Also in such applications any reduction of the spectral interference clatter from diamagnetic species coupled with the broadband coverage of the OFC-FRS technique opens up possibilities for measurements requiring access to multiple transitions simultaneously (e.g. precise multi-line thermometry). 

\begin{acknowledgements}
The authors thank Amir Khodabakhsh for help with operating the DROPO. The work at UmU was supported by the Knut and Alice Wallenberg Foundation (KAW 2015.0159) and the Swedish Foundation for Strategic Research (ICA12-0031). The work at Princeton was supported by the DARPA SCOUT program (W31P4Q161001).   
\end{acknowledgements}

\bibliographystyle{applphysb}

\bibliography{references_APB_OFC_FRS}

\end{document}